\documentclass[a4paper,11pt]{article}
\usepackage{pos}

\usepackage{amsmath,amssymb,physics,bm}
\usepackage{graphicx}
\usepackage{float}
\usepackage{hyperref}
\newcommand{\mA}{\mathcal{A}}

\title{Constrained Symplectic Quantization: Disclosing the Deterministic Framework Behind Quantum Mechanics}
\ShortTitle{Constrained Symplectic Quantization: Quantum Mechanics}

\author*[a,b]{Martina Giachello}
\author[c,d]{Francesco Scardino}
\author[e]{Giacomo Gradenigo}

\affiliation[a]{Gran Sasso Science Institute, Viale F. Crispi 7, 67100 L'Aquila, Italy}
\affiliation[b]{INFN-Laboratori Nazionali del Gran Sasso, Via G. Acitelli 22, 67100 Assergi (AQ), Italy}
\affiliation[e]{Physics and Astronomy Department "Galileo Galilei", Universit\`a di Padova, Via Marzolo 8, 35131 Padova, Italy}
\affiliation[c]{Physics Department, INFN Roma1, Piazzale A. Moro 2, Roma, I-00185, Italy}
\affiliation[d]{Physics Department, Sapienza University, Piazzale A. Moro 2, Roma, I-00185, Italy}

\emailAdd{martina.giachello@gssi.it}
\emailAdd{francesco.scardino@uniroma1.it}
\emailAdd{giacomo.gradenigo@unipd.it}

\abstract{
Symplectic quantization is a functional approach to quantum field theory that allows sampling of quantum fluctuations directly in Minkowski space time by means of a generalized Hamiltonian dynamics in an extra time variable $\tau$ which, at large times, samples a microcanonical ensemble. In a previous work~\cite{Giachello_jhep} we showed that, for an interacting scalar theory in $1+1$ dimensions, this framework captures genuine real time features that are inaccessible to Euclidean simulations. That original formulation, however, suffers from two structural limitations, an ill defined non interacting limit and the lack of a direct correspondence between its correlation functions and those generated by the Feynman path integral.

To solve these problems we introduced constrained symplectic quantization~\cite{Giachello_CSQ1}, a holomorphic reformulation in which fields and action are analytically continued from $\mathbb{R}$ to $\mathbb{C}$ and constraints are imposed on the intrinsic time Hamiltonian flow. The constraints select stable deterministic trajectories already for the free theory and they define convergent holomorphic integration cycles for the corresponding microcanonical measure. In the continuum limit we establish exact equivalence with the Feynman path integral at the level of the generating functional, thus providing a direct link between intrinsic time correlators and real time Green functions.

In this contribution, we apply the method to the quantum harmonic oscillator on a real-time $1$-dimensional lattice. Testing various observables, we find agreement between numerical and exact results for one- and two-point functions, and we reconstruct characteristic real-time features such as an oscillatory propagator, the discrete energy-gap spectrum, and the evolution of eigenstate probability densities. These tests provide numerical evidence that constrained symplectic quantization can sample real-time quantum observables and offers a practical route beyond Euclidean-time importance sampling.
}
\FullConference{The 42nd International Symposium on Lattice Field Theory (LATTICE2025)\\
2--8 November 2025\\
Tata Institute of Fundamental Research, Mumbai, India\\}

\begin{document}
\maketitle

\section{Introduction}
In Euclidean Quantum Field Theory the canonical weight of the Feynman Path integral takes the form $\exp(-S_E/\hbar)$, so the path-integral measure is real and non-negative and the theory can be interpreted as a system in thermal equilibrium, i.e. a \emph{canonical ensemble} at temperature $T$. This simple property is the cornerstone of lattice simulations: observables can be evaluated by importance sampling, and Monte Carlo-based algorithms~\cite{Creutz:1980gp,Parisi:1980ys,Damgaard:1987rr} generate representative field configurations with controlled statistical errors.\par

The situation completely changes in real time. Indeed, in Minkowski signature, quantum amplitudes are weighted by $\exp(iS/\hbar)$ which, being an oscillatory factor, does not define a probability density. As a consequence, importance sampling methods that rely on a positive definite canonical probability distribution lose their foundation~\cite{Gattringer:2016kco,Alexandru:2020wrj}. Developing robust tools for real-time dynamics is essential both for genuine out-of-equilibrium phenomena and for real-time observables that are difficult to reconstruct reliably from Euclidean data. Several approaches have been explored in this direction, including stochastic quantization via complex Langevin dynamics~\cite{Parisi:1983mg} and contour-deformation techniques in complexified field space~\cite{Witten:2010cx} (such as Lefschetz-thimbles formulations~\cite{Cristoforetti:2012su, Alexandru:2020wrj, Scorzato:2015qts, Behtash:2015loa}), each with its own advantages and limitations~\cite{Aarts:2008rr}.\par

Symplectic Quantization (SQ)~\cite{Gradenigo_1,Gradenigo_2,Gradenigo_3} proposes a different route: quantum expectation values are obtained from a \emph{microcanonical} construction generated by a Hamiltonian flow in an auxiliary ``intrinsic time'', denoted as $\tau$. The field operator is promoted from $\hat\phi(x)\to\phi(x,\tau)$ and a new conjugate momentum $\pi(x,\tau)$ is introduced with respect to $\tau$. The intrinsic-time  dynamics is deterministic, Hamiltonian and, by assuming ergodicity, microcanonical expectation values of observables are computed as long-time averages along the trajectory
\begin{equation}
\langle \mathcal{O} \rangle \equiv
\lim_{\Delta\tau\to\infty}\frac{1}{\Delta\tau}
\int_{\tau_0}^{\tau_0+\Delta\tau} d\tau\;
\mathcal{O}\!\left[\phi(\cdot,\tau)\right].
\end{equation}
As said above, the Hamiltonian flow samples a microcanonical measure at fixed generalized action $\mA$,
\begin{equation}\label{eq:microcanonical-prob}
\varrho_{\rm micro}[\phi,\pi] \;=\; \frac{1}{\Omega(\mA)}\,
\delta\!\left(\mathbb{H}[\phi,\pi]-\mA\right),
\qquad
\Omega(\mA)=\int \mathcal{D}\phi\,\mathcal{D}\pi\;
\delta\!\left(\mathbb{H}[\phi,\pi]-\mA\right).
\end{equation}
From an algorithmic viewpoint, this makes SQ close in spirit to Hybrid Monte Carlo (HMC)~\cite{Duane:1987de,Kennedy:2013c,Creutz:1988wv}: in both cases one enlarges the configuration space by introducing momenta and generates proposals via a symplectic integration of Hamilton equations. The difference between SQ and HMC-like methods is both conceptual and practical: in the HMC the Hamiltonian flow is only a \emph{device} to sample a given \emph{canonical} distribution, and an accept/reject step restores the desired canonical probability measure. On the other hand, in SQ the Hamiltonian flow in intrinsic time $\tau$ is instead the \emph{definition} of the sampling procedure, and the expectation values are obtained from long-time averages on a surface of generalized-action. This is precisely why SQ is potentially effective in Minkowski space-time, since the microcanonical construction defines probability as in Eq.~\eqref{eq:microcanonical-prob}, i.e. in a way that does not require the action to be positive definite in order to have a well defined Boltzmann weight.

\section{A first real formulation of Symplectic Quantization}
In the original SQ formulation~\cite{Gradenigo_1,Giachello_jhep} one follows, as closely as possible, the logic that makes Hybrid Monte Carlo effective in the Euclidean case, where the generalized action is taken as 
\begin{equation}\label{eq:HMC}
\mathbb{H}_{\textrm{HMC}}(\phi,\pi)=\frac12\int d^d x\;\pi^2(x,\tau)\;+\;S_E[\phi]\,,
\end{equation}
with a kinetic term depending on the generalized momenta, plus an action-derived potential. The natural Minkowskian analogue is then to 
choose a generalized Hamiltonian whose ``potential'' is directly built from the target Minkowskian action $S$, i.e.\
\begin{equation}\label{eq:origH}
\mathbb{H}_{\mathrm{SQ}}(\phi,\pi)=\frac12\int d^d x\;\pi^2(x,\tau)\;-\;S[\phi]\,,
\end{equation}
so that the intrinsic-time Hamilton equations generate an evolution in $\tau$ controlled by $S[\phi]$. This choice mirrors the Euclidean construction, with the crucial difference that the Minkowski action enters with a sign/structure that is not associated with a positive Boltzmann weight when a classic ensemble change transformation is performed.\par

\begin{figure}[H]
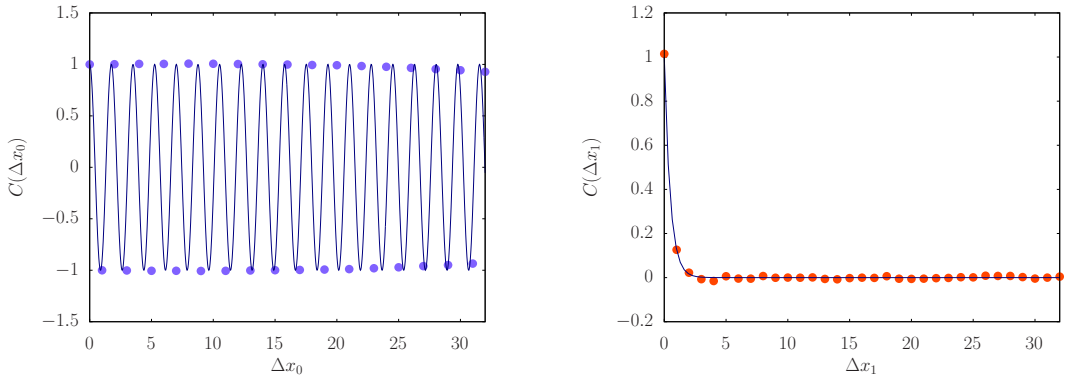

  \begin{minipage}{0.5\textwidth}
    \centering
    \includegraphics[width=0.9\textwidth]{correlationfunction_t_fringe.png}
  \end{minipage}%
  \begin{minipage}{0.5\textwidth}
    \centering
    \includegraphics[width=0.9\textwidth]{correlationfunction_x_fringe.png}
  \end{minipage}
  \caption{Two-point correlation function in real space for a $\lambda\phi^4$ theory in $1+1$ dimensions
  simulated on a square lattice of size $L=128$, lattice spacing $a=1.0$ and mass $m=1.0$. The nonlinearity is set to $\lambda=0.001$).
  \textbf{Left:} oscillations along the $x_0=ct$ (temporal) axis.
  \textbf{Right:} exponential decay along the $x_1$ (spatial) axis, consistent with causal propagation.}
  \label{fig:correlators_SQ}
\end{figure}

In the first lattice tests~\cite{Giachello_jhep,Giachello_lattice2024,Giachello_trani} we considered $\lambda\phi^4$ theory in $1+1$ dimensions with action
\begin{equation}
S[\phi]=\frac12\int d^2x\left\{(\partial_{x_0}\phi)^2-(\partial_{x_1}\phi)^2-m^2\phi^2-\frac12\lambda\phi^4\right\},
\end{equation}
which leads after discretization to intrinsic-time equations of motion of the schematic form
\begin{equation}\label{eq:eom_orig}
\partial^2_\tau{\phi}(x_i,\tau)= -\Delta_0\phi+\Delta_1\phi-m^2\phi(x_i,\tau)-\lambda\,\phi^3(x_i,\tau)\,,
\end{equation}
where $\Delta_0$ and $\Delta_1$ denote the discrete second derivatives in the physical time and space directions, respectively. These tests showed that SQ can reproduce both space-like and time-like correlators in a way consistent with relativistic causality, as illustrated in Fig.~\ref{fig:correlators_SQ}. In particular, when numerically measuring the two point function, one observes exponential decay along the spatial direction and oscillatory behavior along the temporal direction in real time.
The same numerical evidence that validates causality also highlights the central limitations of the original construction. First, the \emph{free} theory is not well-defined: when $\lambda\to 0$ the intrinsic-time evolution generated by Eq. \eqref{eq:origH} becomes unbounded and the microcanonical sampling breaks down, indicating that the interaction plays a stabilizing role in this setup. Secondly, even at small but finite coupling the two-point functions can show an internal inconsistency: the effective mass extracted from the spatial correlator and from the temporal correlator cannot be brought into agreement with the input $m$ (see~\cite{Giachello_jhep}). This discrepancy persists under standard improvements such as larger statistics, smaller step size in the $\tau$ integrator, larger volumes and finer lattices. This is signaling that the problem is not a tunable numerical artifact but a structural limitation of the original formulation~\cite{Giachello_jhep,Giachello_lattice2024,Giachello_trani}. Indeed, analytical calculations (see~\cite{Giachello_jhep,Giachello_lattice2024}) showed that in the large-$N$ degrees of freedom limit (continuum) one does not recover $\int \mathcal{D}\phi\; \exp\!\left(\frac{i}{\hbar}S[\phi]\right)$, but rather a real exponential, schematically
\begin{equation}
\Omega(\mA=\hbar N)\xrightarrow[N\to\infty]{}\int \mathcal{D}\phi\;
\exp\!\left(\frac{S[\phi]}{\hbar}\right).
\end{equation}
In other words, we have a microcanonical sampling associated with a real weight, while we would like the oscillatory Feynman weight. This formula perfectly matches the observation that even-power, bounded potentials stabilize the integral, while the free theory is divergent. Moreover, the absence of the $i$ factor in front of the action makes clear why the mass could not be correctly recovered.
%
%
%
\section{Constrained Symplectic Quantization}\label{sec:CSQ}
The shortcomings of the original real formulation make clear that the problem is not a numerical one, but rather that the \emph{ansatz} for the generalized Hamiltonian must be modified. The key idea of our new formulation~\cite{Giachello_CSQ1}, that takes the name of Constrained Symplectic Quantization (CSQ) is to perform an \emph{analytic continuation} of the degrees of freedom and the action from $\mathbb{R}$ to $\mathbb{C}$ and to build the generalized Hamiltonian out of the \emph{holomorphic} action, in such a way that the associated microcanonical generating functional matches the Feynman path integral in the continuum limit.\par

One first complexifies the field variables $\phi(x,\tau)\in\mathbb{R}\longrightarrow\phi(x,\tau)\in\mathbb{C}$ and $\pi(x,\tau)\in\mathbb{R}\longrightarrow\pi(x,\tau)\in\mathbb{C}$, then the generalized Hamiltonian is chosen as
\begin{equation}\label{eq:H_CSQ}
\mathbb{H}[\phi,\bar{\phi},\pi,\bar{\pi}]
\;=\;\mathbb{K}[\pi,\bar{\pi}] \;+\; 2\,\Im S[\phi,\bar{\phi}]\,,
\qquad
\mathbb{K}[\pi,\bar{\pi}] \equiv \int d^dx\;\bar{\pi}(x,\tau)\,\pi(x,\tau)\,,
\end{equation}
with 
\begin{equation}\label{eq:imS}
\Im S[\phi,\bar \phi]\equiv \frac{S[\phi]-\bar S[\bar \phi]}{2i}\,.
\end{equation}
Notice how the kinetic term is positive and the ``potential'' is given by the imaginary part of the (holomorphically continued) Minkowski action\footnote{Here the imaginary part of the action is a slight abuse of notation since the $\phi$ and $\bar{\phi}$ degrees of freedom are independent off-shell.}. Given Eq.~\eqref{eq:H_CSQ}, if we compute the microcanonical generating functional at generalized action $\mA=\hbar N$ (for the full proof see~\cite{Giachello_CSQ1}) with $N$ the number of degrees of freedom, i.e.
\begin{equation}\label{eq:Omega_CSQ}
\Omega(\mA=\hbar N)
\;=\;
\int \mathcal{D}\bar{\phi}\,\mathcal{D}\phi\,     \mathcal{D}\bar{\pi}\,\mathcal{D}\pi\;
\delta\;\Big(\mathbb{H}[\phi,\bar{\phi},\pi,\bar{\pi}]-\mA\Big)\,,
\end{equation}
the constrained (holomorphic) structure implies that, in the large-$N$ scaling limit corresponding to the continuum limit, the saddle-point evaluation of Eq.~\eqref{eq:Omega_CSQ} yields the desired Minkowski generating functional \cite{Giachello_CSQ1}
\begin{equation}\label{eq:Omega_to_Z}
\Omega(\mA=\hbar N)\xrightarrow[N\to\infty]{}\; Z(\hbar)
\;=\;
\int_{\mathcal{C}}\mathcal{D}\phi\;
\exp\!\left(\frac{i}{\hbar}S[\phi]\right).
\end{equation}
The integration domain $\mathcal{C}$ is a chosen path in the complexified configuration space. In other words, the ``new ansatz'' of Eq. \eqref{eq:H_CSQ} is designed so that the microcanonical formulation matches the \emph{Feynman} weight (and not a real exponential) once the appropriate contour/constraints
are imposed. Even if the complexification apparently doubles variables, CSQ is \emph{not} introducing new degrees of freedom: the Hamiltonian dynamics is constrained to stay on a stable submanifold which corresponds to the choice of a suitable $1$-dimensional integration contour $\mathcal{C}\in \mathbb{C}$ that makes Eq.~\eqref{eq:Omega_to_Z} well defined. Therefore, while the dynamics/integration now happens on an higher dimensional complexified space, the constraints reduce the number of degrees of freedom to match the ones of the original theory.

At this stage it is essential to test the construction in a setting where both stability and correctness can be assessed unambiguously. For this reason we turn to the quantum harmonic oscillator, which provides a clean benchmark as it is exactly solvable. In fact, real-time correlators and the spectrum are known in closed form, allowing direct quantitative checks.
%
%
%
\subsection{Harmonic oscillator: Hamilton equations, mode decomposition and contour rotations}
\label{subsec:CSQ_QHO}
For quantum mechanics we treat the coordinate time as $x_0$ and promote the operator $\hat q(x_0)$ to a
complex field $q(x_0,\tau)\in\mathbb{C}$ fluctuating in the intrinsic time $\tau$,
\begin{equation}
q(x_0,\tau)=q_R(x_0,\tau)+i\,q_I(x_0,\tau),
\qquad
\pi(x_0,\tau)=\pi_R(x_0,\tau)+i\,\pi_I(x_0,\tau),
\end{equation}
together with their anti-holomorphic counterparts $\bar q,\bar\pi$.
The holomorphic oscillator action is kept in its standard form,
\begin{equation}
S[q]=\int_{x_0^i}^{x_0^f}dx_0\left[\frac{m}{2}\left(\frac{dq}{dx_0}\right)^2-\frac12 m\Omega^2 q^2(x_0)\right],
\end{equation}
but now it is evaluated on complex configurations, hence $S[q]\in\mathbb{C}$. The generalized Hamiltonian that defines the intrinsic-time flow is chosen as in Eq.~\eqref{eq:H_CSQ} and Eq.~\eqref{eq:imS}:
\begin{equation}\label{eq:hamiltonian_csq}
\mathbb{H}[\pi,\bar\pi,q,\bar q]=\bar\pi\cdot\pi+2\,\Im S[q,\bar q],
\qquad
\Im S[q,\bar q]\equiv \frac{S[q]-\bar S[\bar q]}{2i},
\end{equation}
which is \emph{real} by construction. The intrinsic-time Hamilton equations are then
\begin{align}\label{eq:hamilton-eqs_csq}
\partial_\tau q=\frac{\partial\mathbb{H}}{\partial\pi}=\bar\pi,
\qquad
\partial_\tau{\bar q}=\frac{\partial\mathbb{H}}{\partial\bar\pi}=\pi,
\qquad
\partial_\tau\pi=-\frac{\partial\mathbb{H}}{\partial q}= i\,\frac{\partial S[q]}{\partial q},
\qquad
\partial_\tau{\bar\pi}=-\frac{\partial\mathbb{H}}{\partial\bar q}= -i\,\frac{\partial \bar S[\bar q]}{\partial\bar q}.
\end{align}
Combining these equations yields a second–order equation that mixes $q$ and $\bar q$,
\begin{equation}\label{eq:complex-eom_csq}
-i\,\frac{\partial^2}{\partial\tau^2}\bar q(x_0,\tau) = - m\,\frac{\partial^2}{\partial x_0^2}q(x_0,\tau)-m\Omega^2 q(x_0,\tau).
\end{equation}
One can then diagonalize the $x_0$-dependence by Fourier decomposition (e.g.\ with periodic boundary conditions for clarity),
\begin{equation}\label{eq:fourier_csq}
q(x_0,\tau)=\sum_{k_0} e^{ik_0 x_0}\,q(k_0,\tau), \qquad k_0=\frac{2\pi\ell}{x_0^f-x_0^i},\ \ \ell\in\mathbb{Z},
\end{equation}
so that each mode satisfies
\begin{equation}\label{eq:harmonic-fourier_csq}
\frac{d^2}{d\tau^2}\bar q(k_0,\tau)+ i\,\omega^2(k_0)\,q(k_0,\tau)=0, \qquad \omega^2(k_0)=m(\Omega^2-k_0^2).
\end{equation}
Writing $q(k_0,\tau)=q_R(k_0,\tau)+i\,q_I(k_0,\tau)$ gives the coupled real system
\begin{align}\label{eq:coupled-real-imag_csq}
\partial^2_\tau\, q_R(k_0,\tau)-\omega^2(k_0)\,q_I(k_0,\tau)=0, \qquad
\partial^2_\tau\, q_I(k_0,\tau)-\omega^2(k_0)\,q_R(k_0,\tau)=0,
\end{align}
which, without further restrictions, admits many unbounded solutions. The \emph{constrained} formulation selects the stable manifold by imposing mode-dependent linear relations between $q_R$ and $q_I$:
\begin{equation}\label{eq:constraints_modes}
q_R(k_0)=-\,q_I(k_0)\quad \text{for }\ \omega^2(k_0)>0
\qquad\text{and}\qquad
q_R(k_0)=+\,q_I(k_0)\quad \text{for }\ \omega^2(k_0)<0.
\end{equation}
Geometrically, Eq.~\eqref{eq:constraints_modes} is equivalent to rotating the integration contour for each Fourier mode by an angle $\pm\pi/4$ in the complex plane:
\begin{equation}\label{eq:rotations_modes}
\omega^2(k_0)>0:\quad q(k_0)=e^{+i\pi/4}\,\tilde q(k_0), \qquad
\omega^2(k_0)<0:\quad q(k_0)=e^{-i\pi/4}\,\tilde q(k_0),
\end{equation}
with $\tilde q(k_0)\in\mathbb{R}$ along the rotated contour. With the rotations of Eq.~\eqref{eq:rotations_modes} the quadratic piece of the action becomes damped mode by mode. Indeed, for a single Fourier mode the holomorphic quadratic contribution can be written (up to an overall constant) as
\begin{equation}\label{eq:feynman_mode}
\exp\!\left(\frac{i}{\hbar}S_k[q]\right)
=\exp\!\left(\frac{i}{2\hbar}\omega^2(k_0)\, q^2(k_0)\right).
\end{equation}
If we now rotate the integration variable as in Eq.~\eqref{eq:rotations_modes},
$q(k_0)=e^{\pm i\pi/4}\tilde q(k_0)$ with $\tilde q(k_0)\in\mathbb{R}$ along the chosen cycle, then
$q^2(k_0)=e^{\pm i\pi/2}\tilde q^2(k_0)=\pm i\,\tilde q^2(k_0)$, and therefore
\begin{equation}\label{eq:damped_gaussian}
\exp\!\left(\frac{i}{2\hbar}\omega^2(k_0)\, q^2(k_0)\right)
=
\exp\!\left(\frac{i}{2\hbar}\omega^2(k_0)(\pm i)\tilde q^2(k_0)\right)
=
\exp\!\left(\mp\,\frac{\omega^2(k_0)}{2\hbar}\,\tilde q^2(k_0)\right).
\end{equation}
The sign in the rotation is chosen such that the exponent is \emph{negative} for that mode, i.e.\ $\mp\,\omega^2(k_0)<0$, producing an ordinary \emph{convergent Gaussian} weight along the rotated contour. Equivalently, the mode weight can be written in the manifestly damped form
\begin{equation}
\exp\!\left(\frac{i}{2\hbar}\omega^2(k_0)\, q^2(k_0)\right)\;\longrightarrow\;
\exp\!\left(-\frac{|\omega^2(k_0)|}{2\hbar}\,\tilde q^2(k_0)\right).
\end{equation}
In the Hamiltonian dynamics this same choice removes the runaway directions of the coupled system of Eq.~\eqref{eq:coupled-real-imag_csq}: after imposing the linear constraint (equivalently the $\pm\pi/4$ rotation), each mode reduces to stable harmonic motion in intrinsic time $\tau$ with bounded trajectories, so the microcanonical evolution is well-defined.
%
%
%
%
\subsection{Numerical tests on the lattice}\label{subsec:numerics}
We now want to to test the dynamics of Eq.~\eqref{eq:complex-eom_csq} with constraints of Eq.~\eqref{eq:constraints_modes} on the lattice, so we discretize the physical time direction as $x_0^{(\ell)}=\ell a$ (with $\ell=0,\dots,N_t-1$, $T=N_t a$) and evolve the constrained Hamilton equations in the intrinsic time $\tau$ with a symplectic integrator. Observables are computed as long-$\tau$ averages after thermalization. We report results for three complementary setups: periodic boundaries, fixed--fixed boundaries, and fixed+free boundaries.
\subsubsection{Two-point function}
\begin{figure}[H]
  \centering
  \includegraphics[width=0.95\textwidth]{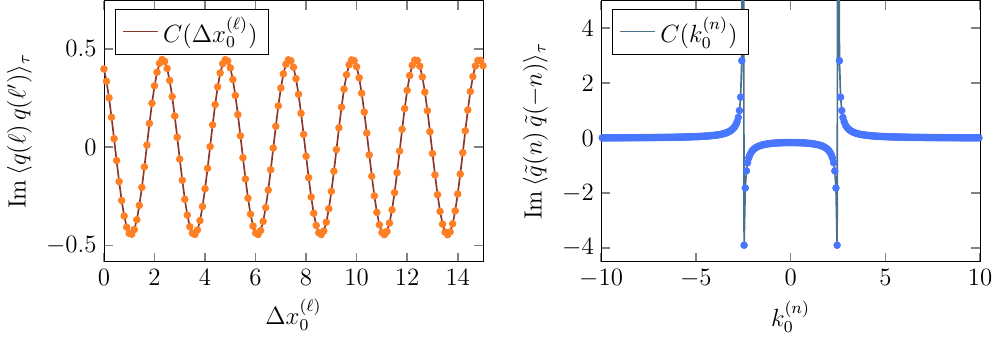}
  \caption{\textbf{Left:} coordinate-space correlator $\Im\,\langle q(\ell)\,q(\ell')\rangle_{\tau}$
    compared with the analytic lattice prediction of Eq.~\eqref{eq:2pt-ho-lattice-PBC}.
    \textbf{Right:} Fourier-spectrum correlator $\Im\,\langle \tilde q(n)\,\tilde q(-n)\rangle_{\tau}$
    compared with the prediction of Eq.~\eqref{eq:2pt-ho-momentum-lattice}.
    Simulation parameters: $N_t=1024$, $a=0.1$, $m=1$, $\Omega=2.5$, $d\tau=0.01$, periodic boundaries.}
  \label{fig:qho_corr_both}
\end{figure}

We impose periodic boundary conditions in physical time, $x_0^{(\ell)}=\ell a$ with $\ell=0,\dots,N_t-1$ and $T\equiv N_t a$, and measure the translation-invariant correlator as an intrinsic-time average
\begin{equation}
\Im\langle q(\ell)\,q(\ell')\rangle_\tau,
\qquad
\langle \mathcal{O}\rangle_\tau=\lim_{\Delta\tau\to\infty}\frac{1}{\Delta\tau}
\int_{\tau_{\rm eq}}^{\tau_{\rm eq}+\Delta\tau}\!d\tau\;\mathcal{O}(\tau).
\end{equation}
For comparisons we use the exact free lattice prediction for the harmonic oscillator with periodic boundaries,
\begin{align}
  C(\Delta x_0^{(\ell)})\equiv\langle q(\ell)\, q(\ell') \rangle_{\text{P.B.}}^{\text{(lat)}}
  = -\frac{i\,a}{2m\sin(\kappa a)}\,
  \frac{\sin\!\big[\kappa\,a(\ell-\ell')\big]
        + \sin\!\big[\kappa\,(T-a(\ell-\ell'))\big]}
       {1-\cos(\kappa T)} ,
  \label{eq:2pt-ho-lattice-PBC}
\end{align}
with $\kappa$ fixed by the lattice dispersion relation, $\cos(\kappa a)=1-\frac{a^2\Omega^2}{2}$. In momentum space we compute
\begin{equation}
\Im\langle \tilde q(n)\tilde q(-n)\rangle_\tau,
\quad\textrm{with}\quad
\tilde q(n)=\frac{1}{\sqrt{N_t}}\sum_{\ell=0}^{N_t-1}e^{-i k_0^{(n)} \ell a}\,q(\ell)
\qquad
k_0^{(n)}=\frac{2\pi n}{T},
\end{equation}
and compare with the corresponding free prediction in momentum space,
\begin{align}
 C(k_0^{(n)})\equiv \Big\langle \tilde q\!\left(n\right)\,\tilde q\!\left(-n\right)\Big\rangle_{\text{P.B.}}^{\text{(lat)}}
  = \frac{i}{m}\,
  \frac{1}{\frac{4}{a^2}\sin^2\!\left(k_0^{(n)} a/2\right)-\Omega^2}\,,
  \label{eq:2pt-ho-momentum-lattice}
\end{align}
which follows from the discrete kernel
$\hat k_0^{\,2}(n)=\frac{4}{a^2}\sin^2(k_0^{(n)}a/2)$ on the periodic lattice.
Figure~\ref{fig:qho_corr_both} shows the measured correlators in coordinate and momentum space together with
the theoretical curves; agreement provides a direct check that the constrained dynamics reproduces the
free real-time propagator on the periodic lattice.
%
%
%
%
\subsubsection{Energy spectrum}
\label{sec:energyeigenvalues}
\begin{figure}[H]
  \centering  \includegraphics[width=1\textwidth]{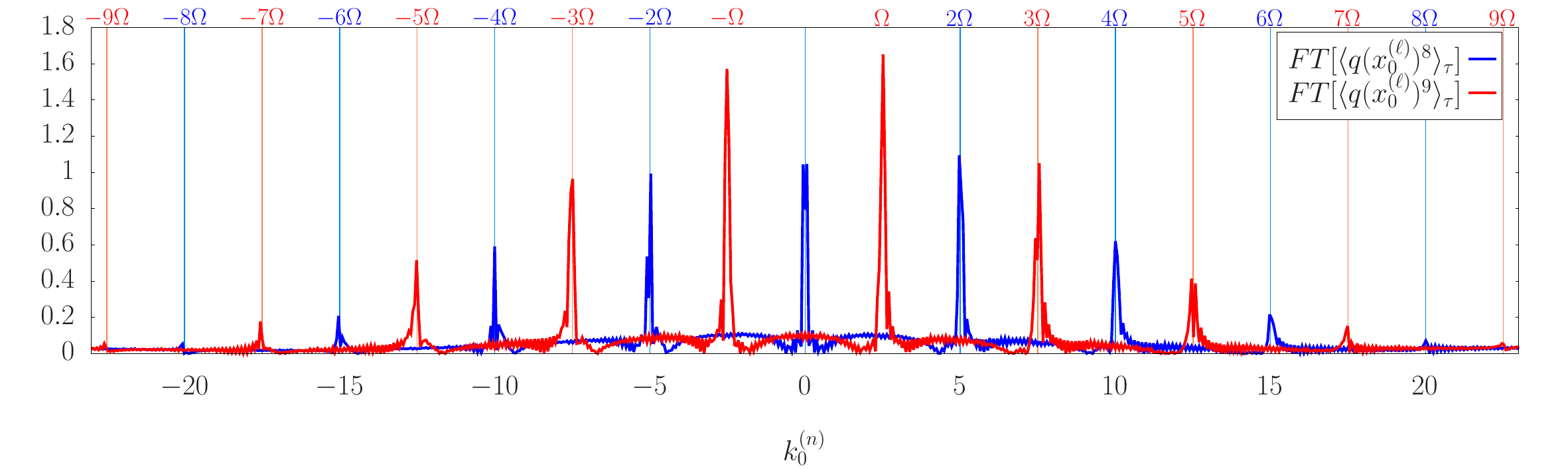}
  \caption{Fourier spectra of fixed--fixed signals (here $q^j$ with $j=8,9$) for the harmonic oscillator.
  Simulation parameters: $m=1.0$, $\Omega=2.5$, $a=0.1$, $d\tau=10^{-2}$,
  $N_t=1024$; fixed boundaries $q(x_0^i)=1.021$, $q(x_0^f)=-2.414\times10^{-1}$.}
  \label{fig:eigenvalues_summary}
\end{figure}

Differently from the Euclidean setting, where imaginary-time evolution exponentially suppresses excited-states contributions and effectively projects onto the ground state, the real-time formulation preserves coherent oscillations. As a result, with fixed–fixed boundary conditions the matrix elements are generically time-dependent and their Fourier spectra directly resolves the energy gaps, enabling a direct reconstruction of the energy spectrum. We impose boundary conditions in physical time $q(x_0^i)=q_i$ and $q(x_0^f)=q_f$, (with $q_i\neq q_f$ in general) and study the time series of a generic operator insertion $B(x_0)$ between the fixed boundary states, $\langle q_f,x_0^f|\,B(x_0)\,|q_i,x_0^i\rangle$. Inserting complete sets of energy eigenstates gives
\begin{equation}\label{eq:B_time_dep_fixed}
\langle q_f,x_0^f|\,B(x_0)\,|q_i,x_0^i\rangle=\sum_{n,m} b_n^\ast\,c_m\,
\langle \psi_n|B|\psi_m\rangle\,
e^{-i(E_m-E_n)(x_0-x_0^i)/\hbar},
\end{equation}
with coefficients $c_m=\langle \psi_m|q_i\rangle$ and $b_n= e^{-iE_n(x_0^f-x_0^i)/\hbar}\,\langle \psi_n|q_f\rangle,$. The Fourier transform of $\langle q_f,x_0^f|\,B(x_0)\,|q_i,x_0^i\rangle$ therefore exhibits peaks at the energy gaps $\omega_{mn}=(E_m-E_n)/\hbar$.
For the harmonic oscillator $E_n=\hbar\Omega(n+\tfrac12)$, the spectrum consists of equally spaced gaps, so the peaks must occur at integer multiples of $\Omega$. In practice we choose $B(x_0)=q^j(x_0)=(\frac{\hbar}{2m\Omega})^{j/2}(a+a^\dagger)^j$, which provides simple and robust selection rules. Since $q^j(x_0)$ is parity even (odd) for even (odd) $j$, only transitions between states of the same (opposite) parity contribute, and the frequency content is restricted to
\begin{equation}\label{eq:parity-selection-fixed}
j\ \text{even}:\ \omega=r\,\Omega\ \ (r=0,2,4,\dots),
\qquad
j\ \text{odd}:\ \omega=r\,\Omega\ \ (r=1,3,5,\dots).
\end{equation}
Figure~\ref{fig:eigenvalues_summary} shows Fourier spectra of $\langle q^j\rangle$ for $j=8,9$: the observed peaks follow the pattern \eqref{eq:parity-selection-fixed} and are located at the expected multiples of $\Omega$, providing a direct reconstruction of the harmonic-oscillator energy gaps from CSQ data.
%
%
%
%
\subsubsection{Time evolution of the wave-function}
\label{sec:pdf_reco}
\begin{figure}[H]
  \centering
  \makebox[\textwidth][c]{%
    \includegraphics[width=0.35\textwidth]{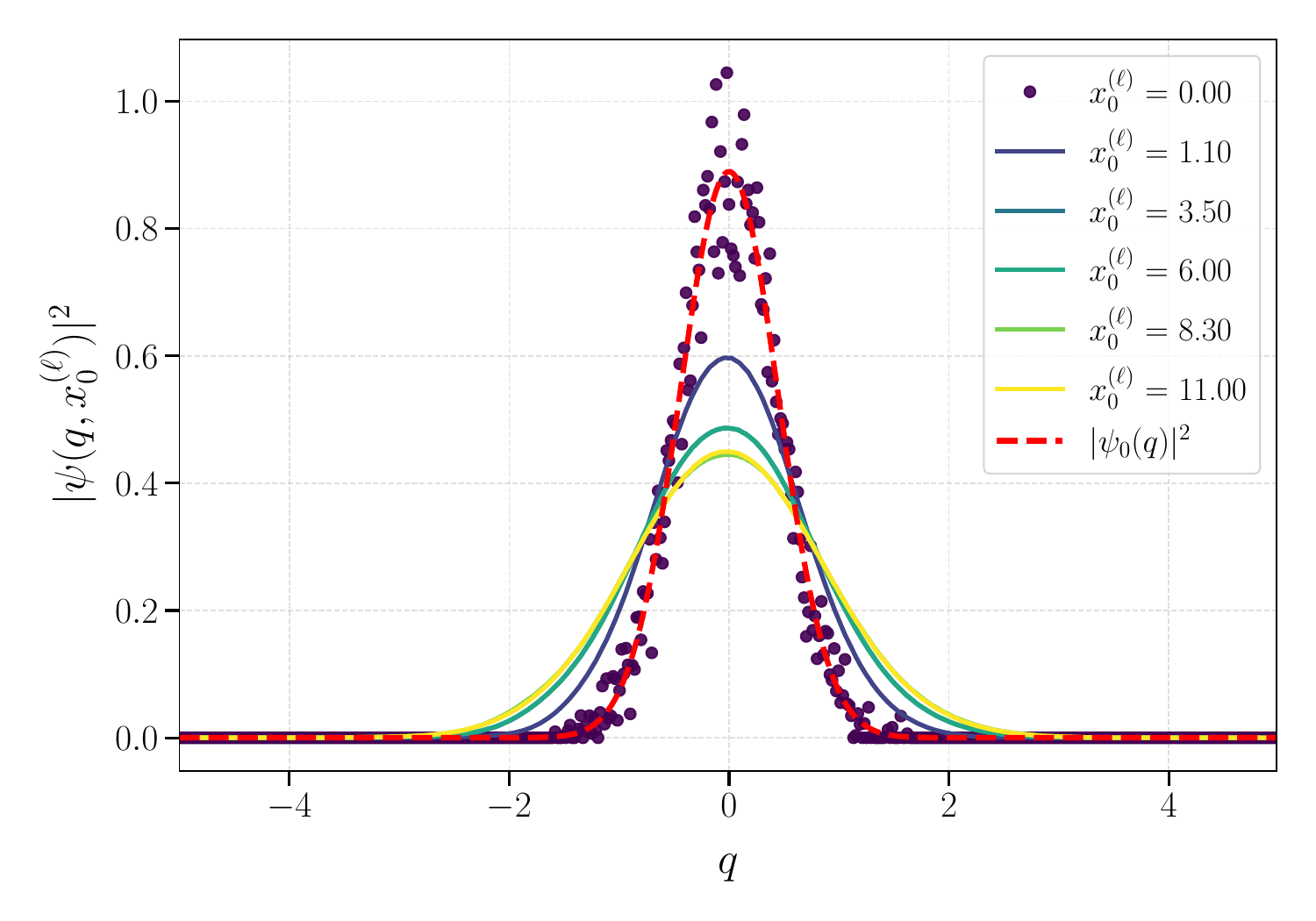}\hspace{0.01\textwidth}%
    \includegraphics[width=0.35\textwidth]{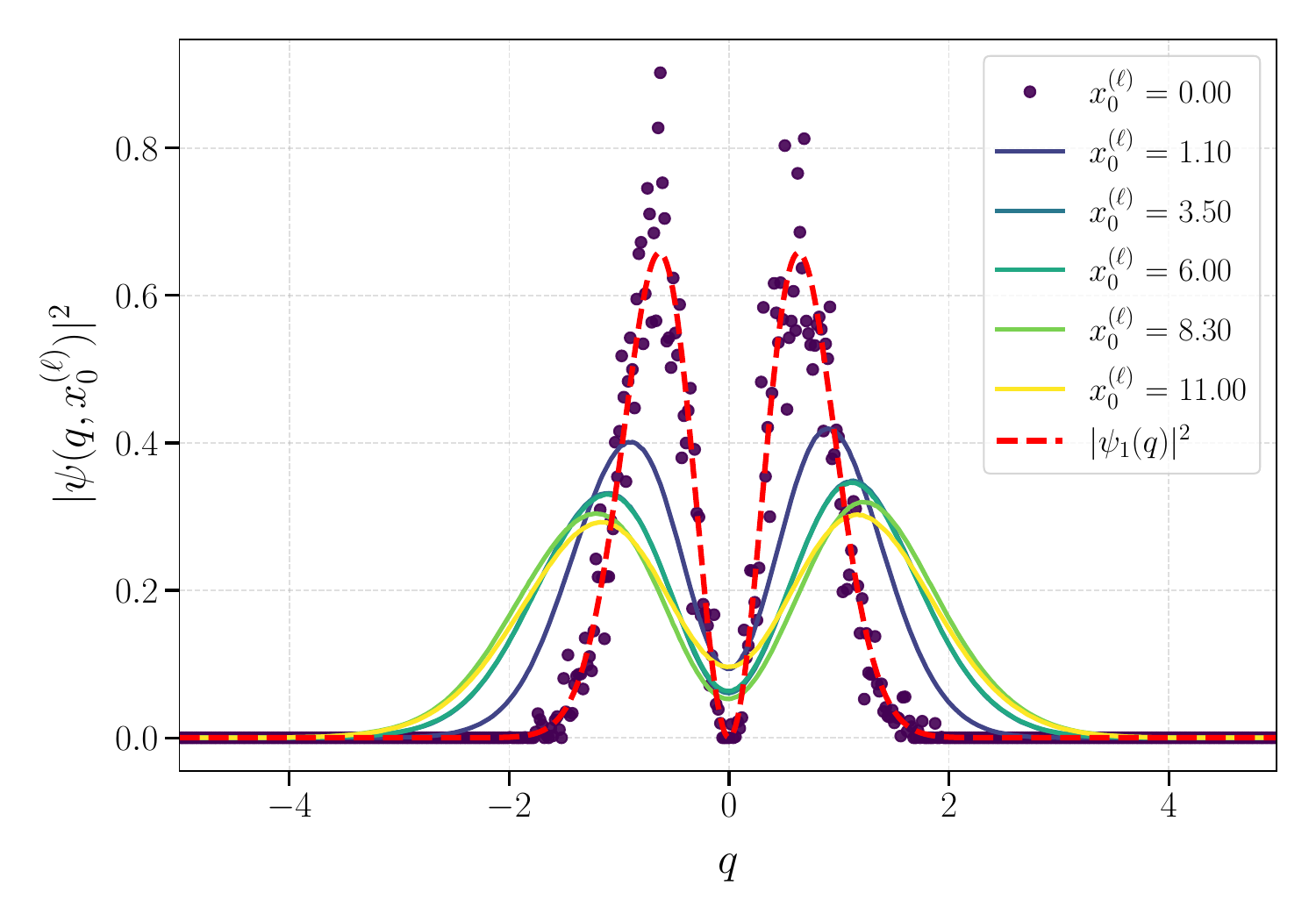}\hspace{0.01\textwidth}%
    \includegraphics[width=0.35\textwidth]{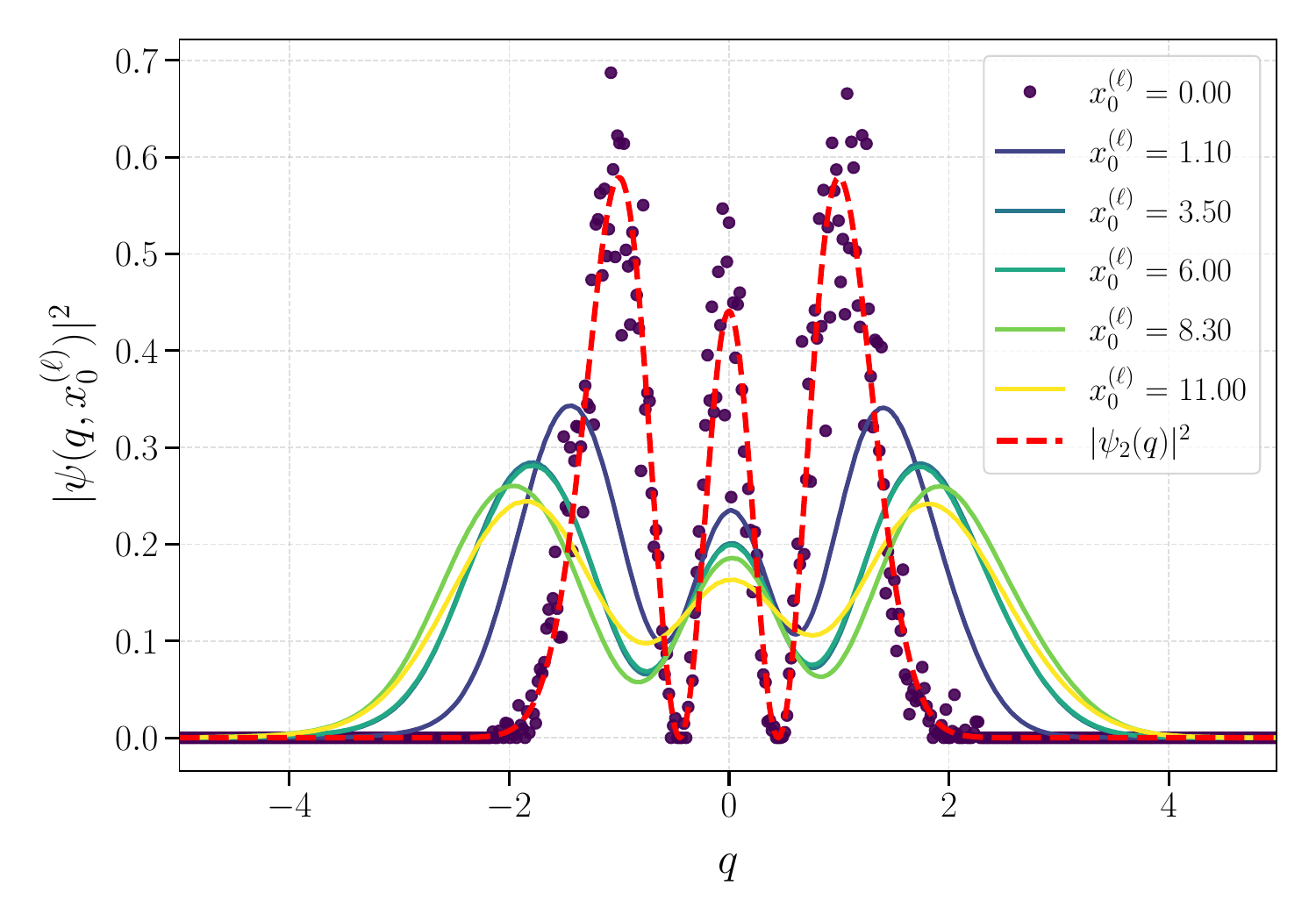}%
  }
  \caption{Eigenstate probability-density reconstruction with fixed+free boundaries.
  Initial conditions are sampled from $P_n(q_0)=|\psi_n(q_0)|^2$ and histograms $P(q,x_0^{(\ell)})$ are shown at
  several intermediate times $x_0^{(\ell)}$. Dashed curves: analytic $|\psi_n(q)|^2$. Left to right: $n=0,1,2$.}
  \label{fig:pdf_reco_states}
\end{figure}

To test CSQ beyond two-point functions we study the real-time evolution of a probability density in a fixed+free setup. We fix the coordinate at the initial time $q(x_0^i)=q_0$ and leave the final endpoint $q(x_0^f)$ unconstrained, and generate an ensemble of trajectories by sampling the initial condition $q_0$ from a prescribed
distribution $P_n(q_0)$. For each draw $q_0^{(r)}$ we evolve the constrained dynamics and record the value $q^{(r)}(x_0^{(\ell)})$ at an intermediate time $x_0^{(\ell)}$. Histogramming these values over the ensemble $r=1,\dots,N_{\rm traj}$ yields an estimator of the probability density at time $x_0^{(\ell)}$,
\begin{equation}
P(q,x_0^{(\ell)})\;\simeq\;\frac{1}{N_{\rm traj}}\sum_{r=1}^{N_{\rm traj}}
\delta\,\big(q-q^{(r)}(x_0^{(\ell)})\big),
\end{equation}
implemented numerically with finite bins, as shown in Fig.~\ref{fig:bc_fixedfree}. In our specific setup we choose as input an energy-eigenstate density, $P_n(q_0)=|\psi_n(q_0)|^2$. Since $|\psi_n(q,x_0)|^2$ is stationary, the exact prediction is $P(q,x_0)=|\psi_n(q,x_0)|^2=|\psi_n(q)|^2$,
so the reconstructed histograms $P(q,x_0^{(\ell)})$ should coincide for all $x_0^{(\ell)}$, even though individual trajectories are non-trivial and oscillatory. This provides a direct check that CSQ reproduces the correct real-time evolution at the level of probabilities.
\begingroup
\setlength{\intextsep}{2pt}      
\setlength{\abovecaptionskip}{0pt}
\setlength{\belowcaptionskip}{0pt}
\begin{figure}[H]
  \centering
  \includegraphics[width=1\textwidth]{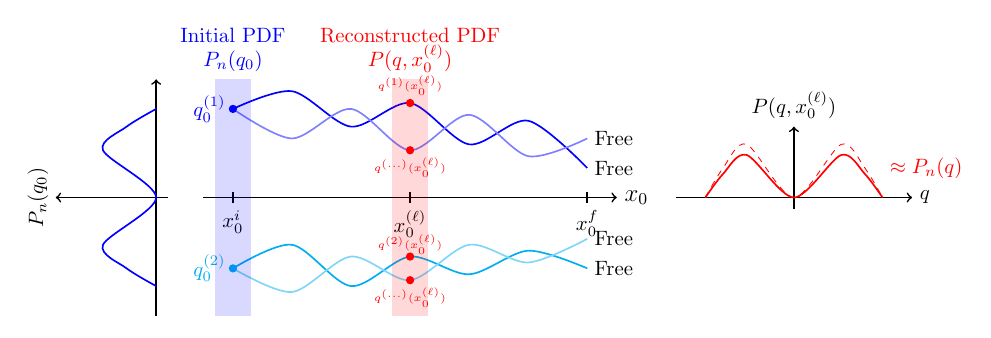}
  \caption{Fixed+free setup used for probability-density reconstruction: an ensemble of initial conditions $q_0^{(r)}$ is drawn from a chosen $P_n(q_0)$ and evolved to
  reconstruct $P(q,x_0^{(\ell)})$ by histogramming.}
  \label{fig:bc_fixedfree}
\end{figure}
\endgroup
Figure~\ref{fig:pdf_reco_states} shows the reconstruction for $n=0,1,2$: points are the CSQ histograms at several $x_0^{(\ell)}$, while the dashed curve is the analytic $|\psi_n(q)|^2$. The agreement is not exact because we initialize an \emph{ensemble density}, sampling $q_0\sim P_n(q_0)=|\psi_n(q_0)|^2$, rather than a coherent \emph{amplitude} with a fixed phase. The reconstruction therefore targets $|\psi(q,x_0)|^2$ (not the full wave-function), and at finite $N_{\rm traj}$ and finite bin width small deviations are most visible near nodes and in the tails. Overall, the fixed+free test shows that CSQ transports probability densities consistently in real time: $P(q,x_0^{(\ell)})$ remains compatible with the stationary eigenstate density for $n=0,1,2$.
\begingroup
\small
\renewcommand{\baselinestretch}{0.98}\selectfont
\bibliographystyle{JHEP}
\bibliography{references}
\endgroup

\end{document}